\documentclass[12pt]{article}

\usepackage{ulem}
\usepackage{epsfig, graphicx}
\usepackage{amsmath}
\usepackage{latexsym}
\usepackage{amstext}
\usepackage{epsfig, graphicx}
\usepackage{epsfig}
\usepackage{latexsym}
\usepackage{amstext}
\usepackage{amssymb}
\usepackage{graphics}
\usepackage{color}

\newcommand{\p}{\partial}
\newcommand{\ds}{\displaystyle}
\newcommand{\beq}{\begin{eqnarray}}
\newcommand{\beqq}{\begin{eqnarray*}}
\newcommand{\eeq}{\end{eqnarray}}
\newcommand{\eeqq}{\end{eqnarray*}}

\newcommand{\x}{\mbox{\boldmath$x$}}
\newcommand{\Aa}{\mbox{\boldmath$A$}}

\newcommand{\As}{\mbox{\boldmath$a$}}
\newcommand{\y}{\mbox{\boldmath$y$}}

\newcommand{\J}{\mbox{\boldmath$J$}}
\newcommand{\ET}{\mbox{\boldmath$\eta$}}

\newcommand{\X}{\mbox{\boldmath$X$}}
\newcommand{\Y}{\mbox{\boldmath$Y$}}
\newcommand{\Yy}{\mbox{\boldmath$y$}}
\newcommand{\Z}{\mbox{\boldmath$Z$}}
\newcommand{\w}{\mbox{\boldmath$w$}}
\newcommand{\vv}{\mbox{\boldmath$v$}}
\newcommand{\bb}{\mbox{\boldmath$b$}}
\newcommand{\Bb}{\mbox{\boldmath$b$}}
\newcommand{\B}{\mbox{\boldmath$B$}}

\newcommand{\aaa}{\mbox{\boldmath$a$}}

\newcommand{\SSigma}{\mbox{\boldmath$\Sigma$}}

\newcommand{\IIm}{\mbox{\boldmath$I_m$}}

\newcommand{\thet}{\mbox{\boldmath$\theta$}}
\newcommand{\intR}{\int\limits_{\mathbb{R}}}
\newcommand{\intRm}{\int\limits_{\mathbb{R}^m}}
\newcommand\norm[1]{\left\lVert#1\right\rVert}
\definecolor{red}{rgb}{1,0,0}

\font\bb=msbm10 at 12pt 
 
\def\rR{\hbox{\bb R}}

\def\ds#1{\displaystyle{#1}}

\def\eE{\hbox{\bb E}}
\begin{document}
\title{Recovering a stochastic process from noisy ensembles of many single particle trajectories}
\author{N. Hoze and D. Holcman \footnote{Ecole Normale Sup\'erieure, 46 rue d'Ulm 75005 Paris, France. Institut f\"ur Integrative Biologie, ETH, Universit\"atstrasse 16 8092 Z\" urich, Switzerland.
}}
\date{\today}
\maketitle
\begin{abstract}
Recovering a stochastic process from noisy ensembles of single particle trajectories (SPTs) is resolved here using the Langevin equation as a model. The massive redundancy contained in SPTs data allows recovering local parameters of the underlying physical model. We use several parametric and non-parametric estimators to compute the first and second moment of the process and to recover the local drift, its derivative and the diffusion tensor.  Using a local asymptotic expansion of the estimators and computing the empirical transition probability function, we develop here a method to deconvolve the instrumental from the physical noise.  We use numerical simulations to explore the range of validity for the estimators. The present analysis allows characterizing what can exactly be recovered from the statistics of super-resolution microscopy trajectories used in molecular trafficking and underlying cellular function.
\end{abstract}

\section{Introduction}
The redundancy of many short single particle trajectories (SPTs) are necessary to extract physical parameters from empirical data at a molecular level \cite{DSP,HozePNAS,HozeBJ2014}, while long isolated trajectories have been used to extract second order properties of a Brownian motion using mean-square displacement analysis \cite{Saxton-97,Saxton2008} \cite{Heinrich1,Heinrich6}. Some geometrical properties can also be recovered from long trajectories, such as the radius of confinement for a confined Brownian motion \cite{Kusumi93}. In the context of cellular transport (ameoboid), high resolution motion analysis of long SPTs \cite{Heinrich2} in micro-fluidic chambers containing obstacles revealed different type of cell motions. Depending on the obstacle density, crawling was found at low density of obstacles \cite{Heinrich3} and directed and random phases can even be differentiated. In high density regions, the motion is rather directed from obstacle-to-obstacle \cite{Heinrich5}.

Under additional assumptions about the physical process and with the advent of massive high resolution microscopy data, it has been recently possible to recover additional features from many short trajectories such as the local drift, the diffusion tensor and even potential wells that are refined local structures, generating confinement due to a direct field of forces \cite{HozePNAS,HozeBJ2014,HolcmanCIB2013}. Moreover, with a model of obstacles organization,  the map of diffusion coefficient can be converted into a density of obstacles \cite{Hoze2011}. Several approaches have been proposed to reconstruct diffusion properties from empirical estimators of a large ensemble of single noisy trajectories \cite{Berglund,Verstergaard}, even when trajectories are sampled and recorded points contain additional noise due to background limitations \cite{Thompson}. Precise and careful estimates \cite{Berglund,Verstergaard} have been obtained for pure diffusion processes (no drift).

In this article, we present a general analysis of short stochastic trajectories that can be driven by a non-constant drift. The drift analysis is relevant when a tracked particle experiences direct interactions or becomes confined by a potential well, that needs to be resolved and parameters are extracted from data. Because empirical data can be potentially noisy, the drift term can be affected by noise in the measurement, such as tracking noise, thus requiring a careful interpretation of the data analysis. As we shall see here, when a stochastic particle crosses a potential well, the second derivative of the potential well is an additional term that contributes to the expression of the measured diffusion coefficient. Thus, a deconvolution of the trajectories is needed to remove instrumental noise or tracking error that affect the recovery of the physical from the measured trajectory.

Deriving analytical formula allows resolving precisely the contribution of each term and recovering the physical dynamics by computing the first and second moments from data. The empirical data are presented as a collection of discrete trajectories obtained at a fixed time resolution $\Delta t$.  We present both parametric and non-parametric estimators and the underlying physical model here is the Smoluchowski limit of Langevin's equation. In addition to several estimators and their analysis, numerical simulations are used to explore the range of validity of the estimators. This article is organized as follows: the first part is dedicated to construct non-parametric empirical estimators from a stochastic analysis in the entire space $\rR^d$, $d=1...n$. Second, we derive analytical formula for the first and the second moment. In the third part, we study the parametric estimators for an Ornstein-Uhlenbeck process and obtain specific estimates. The analytical formula for the estimators are compared to numerical simulations. We conclude that this analysis supports that biophysical properties of a membrane can be recovered from the empirical estimators of many SPTs and potential wells are physical objects \cite{HozePNAS,HozeBJ2014} and not a fiction of tracking algorithms.

\section{Estimations of a stochastic process using a non-parametric estimators}
\subsection{Stochastic Model}
The physical motion of a stochastic particle is modeled by the Smoluchowski's limit of the Langevin equation resulting in the equation of motion 
\beq \label{eqfdt1}
\dot{\X}=\As(\X) + \Bb(\X)\dot{\w},
\eeq
where $\As$ is a deterministic drift, $\Bb$ the diffusion tensor and $\w$ the classical Wiener $\delta$-correlated noise. The Ito's integral leads to
\beq
\X(t) =\X(u) +\int_{u}^{ t}\As(\X(s))ds +\int_{u}^{ t}\Bb(\X(s))d\w_s
\eeq
and at times $0,\Delta t,\dots, n\Delta t$,
\beq
\int_{n\Delta t}^{ (n+1)\Delta t}\As(X(s))ds &=& \As(\X_n)  \Delta t +o(\Delta t)\\
\int_{n\Delta t}^{ (n+1)\Delta t}\Bb(\X(s))d\w_s &=& \Bb(\X_n)\Delta w,
\eeq
the discrete approximation sequence is
\beq \label{eqsyst}
\X_{n+1} =\X_{n} +\As(\X_n) \Delta t  + \Bb(\X_n)\Delta w,
\eeq
where $\X_n=\X(n\Delta t)$. 

We assume that the physical motion is sampled at points $\X_n$: at each time step, an additive Gaussian noise is added and the observed motion is described by
\beq\label{observedp}
\Y_n= \X_n+\Z_n, \mbox{ where } \Z_n = \sigma {\ET_n}
\eeq
and ${\ET_n}$ is a $n$-dimensional Gaussian variable. We shall present various statistical parametric and non-parametric approaches to recover the underlying stochastic component of the continuous variable $\X$ from the empirical measured sequence $\Y_n$.

\subsection{Recovering the empirical transition probability density function in $\rR$}\label{ptf}
We compute here the transition probability $p(\y,n+1|\x,n)=Pr\{\Y_{n+1}=\y|\Y_n=\x\}$ in one dimension when the diffusion tensor $ b(X_n) =\sqrt{2D}$ is uniform:
\beq
p(Y_{n+1}=y|Y_n=x) &=& p(X_{n+1}+Z_{n+1}=y|X_n+Z_n=x).
\eeq
The two processes $X_n$ and $Z_n$ are independent and in $\rR$, we have
\beq
p(Y_{n+1}=y|Y_n=x) &=&  \intR p(X_{n+1}+Z_{n+1}=y|X_n=x_1)p(Z_n=x-x_1)d x_1\nonumber\\
&=& \intR \intR p(X_{n+1}=y_1,Z_{n+1}=y-y_1|X_n=x_1)p(Z_n=x-x_1)dx_1dy_1\nonumber\\
&=&  \intR  \intR p(X_{n+1}=y_1|X_n=x_1)p(Z_{n+1}=y-y_1)p(Z_n=x-x_1)dx_1dy_1 \nonumber\\
&=&  \intR  \intR p(X_{n+1}=y_1|X_n=x_1). \frac{e^{-\ds{\frac{(x-x_1)^2}{2\sigma^2}}}}{\sigma\sqrt{2\pi}}
\frac{e^{-\ds{\frac{(y-y_1)^2}{2\sigma^2}}}}{\sigma\sqrt{2\pi}}dx_1dy_1. \nonumber\\
&& \label{transitionprobabilityintegral}
\eeq
For $\Delta t\ll1$ and $X_{n+1}-X_n \sim \mathcal{N}(a(X_n)\Delta t,\sqrt{2D\Delta t})$, the pdf is
\beq
p(X_{n+1}=y_1|X_n=x_1)=\frac{e^{-\ds{\frac{(y_1-x_1-a(x_1)\Delta t)^2}{4D\Delta t}}}}{\sqrt{4\pi D \Delta t}},
\eeq
which gives that
\beqq
p(Y_{n+1}=y|Y_n=x)  &=&  \intR  \intR  \frac{e^{-\ds{\frac{(y_1-x_1-a(x_1)\Delta t)^2}{4D\Delta t}}}}{\sqrt{4\pi D \Delta t}} \frac{e^{-\ds{\frac{(x-x_1)^2}{2\sigma^2}}}}{\sigma\sqrt{2\pi}}\frac{e^{-\ds{\frac{(y-y_1)^2}{2\sigma^2}}}}{\sigma\sqrt{2\pi}}dx_1dy_1\\
&=& \intR  \frac{e^{-\ds{\frac{(x-x_1)^2}{2\sigma^2}}}}{\sigma\sqrt{2\pi}}  \frac{e^{-\ds{\frac{(y-x_1-a(x_1)\Delta t)^2}{2(\sigma^2+ 2D\Delta t)}}}}{ \sqrt{2\pi(\sigma^2 +2D\Delta t )}} dx_1.
\eeqq
To obtain an explicit expression of this convolution, we use the change of variable $x_1=x+\sigma \eta$ where $\sigma \ll1$,
 \beqq
p(Y_{n+1}=y|Y_n=x) =\intR  \frac{e^{-\ds{\frac{\eta^2}{2}}}}{\sqrt{2\pi}}  \frac{e^{-\ds{\frac{(y-x-\sigma \eta-a(x+\sigma \eta)\Delta t)^2}{2(\sigma^2+ 2D\Delta t)}}}}{ \sqrt{2\pi(\sigma^2 +2D\Delta t )}} d\eta.
\eeqq
Using a Taylor's expansion, we have $ a(x+\sigma \eta) = a(x) + \sigma \eta a'(x) +o(\sigma)$ and 
\beqq
p(Y_{n+1}=y|Y_n=x) &=&  \intR   \frac{e^{-\ds{\frac{\eta^2}{2}}}}{\sqrt{2\pi}} \frac{e^{-\ds{\frac{(y-x-a(x)\Delta t - \sigma \eta (1+a'(x)\Delta t))^2}{2(\sigma^2+ 2D\Delta t)}}}}{\sqrt{2\pi(\sigma^2 +2D\Delta t )}} d\eta \\ & & \\
&=&\ds{\frac{1}{\sigma(1+a'(x)\Delta t)}\intR  \frac{e^{-\ds{\frac{\eta^2}{2}}}}{\sqrt{2\pi}} \frac{e^{-\ds{\frac{(\eta-\frac{y-x-a(x)\Delta t}{\sigma(1+a'(x)\Delta t)})^2}{2\frac{\sigma^2+2D\Delta t}{\sigma^2(1+a'(x)\Delta t)^2}  }}}}{\sqrt{\frac{\sigma^2+2D\Delta t}{\sigma^2(1+a'(x)\Delta t)^2}}\sqrt{2\pi}} d\eta} \\& &\\
&=& \frac{1}{\sigma(1+a'(x)\Delta t)} \frac{{e^{\ds{-\frac{\left(\frac{y-x-a(x)\Delta t}{\sigma(1+a'(x)\Delta t)}  \right)^2}{2\left(1+\frac{\sigma^2+2D\Delta t}{\sigma^2(1+a'(x)\Delta t)^2} \right)}}}}}{\sqrt{2\pi}\sqrt{1+\frac{\sigma^2+2D\Delta t}{\sigma^2(1+a'(x)\Delta t)^2}}} \\ &&\\
&=& \frac{e^{\ds{-\frac{\left(y-x-a(x)\Delta t  \right)^2}{2\left(2\sigma^2(1+a'(x)\Delta t)+2D\Delta t+O(\Delta t)^2\right)}}}}{\sqrt{2\pi}\sqrt{2\sigma^2(1+a'(x)\Delta t)+2D\Delta t+O(\Delta t)^2 }}.
\eeqq
We obtain that
\beq \label{pdfosb}
p(Y_{n+1}=y|Y_n=x) = \frac{e^{-\ds{\frac{\left(y-x-a(x)\Delta t  \right)^2}{2\sigma_{\Delta t}^2(x)}}}}{\sigma_{\Delta t}(x)\sqrt{2\pi}},
\eeq
where
\beq
\sigma_{\Delta t}^2(x)= 2\sigma^2(1+a'(x)\Delta t)+2D\Delta t+O(\Delta t)^2.
\label{sigma1}
\eeq
We conclude that the transition probability of the observed process $Y_n$ is Gaussian and $Y_{n+1}-Y_n \sim \mathcal{N}(a(Y_n)\Delta t,\sigma_1(Y_n))$.  The observed motion is thus defined by the discrete scheme:
\beq\label{empi}
\tilde Y_{\Delta t}(t+\Delta t) &=& \tilde Y_{\Delta t}(t)+ a_{obs}(\tilde Y_{\Delta t}) \Delta t  +\frac{ \sigma_{obs, \Delta t}(\tilde Y_{\Delta t})}{\sqrt{\Delta t}}\Delta W_t,
\eeq
where $\Delta W_t=W(t+\Delta t)-W(t)$ and $W$ is a  Brownian motion of variance 1 and
\beq
a_{obs}(x) &=& a(x) \\
\sigma_{obs, \Delta t}(x) &=& \sigma_{\Delta t}(x)=\sqrt{2\sigma^2(1+a'(x)\Delta t)+2D\Delta t+O(\Delta t)^2}.
\eeq
This approach allows defining the continuous process $\tilde Y_{\Delta t}$ from the approximation at the scale $\Delta t$, it is solution of the stochastic equation
\beq
d\tilde Y_{\Delta t}(s) &=& a(\tilde Y_{\Delta t}) ds  +\frac{ \sigma_{\Delta s}(\tilde Y_{\Delta t})}{\sqrt{\Delta t}}dW_s.
\label{SDEy}
\eeq
The drift of the observed process at a time resolution $\Delta t$ is the same (at first order in $\sigma$) as the physical one, while the diffusion tensor is changed and given by formula \eqref{sigma1}.

\section{Estimating the drift and diffusion tensor}
Optimal estimators of the physical process \eqref{eqfdt1} are constructed from Feller's formula  \cite{HozeBJ2014,OPT, book,DSP}
 \beq
 \label{eq:force}
\As(\X)=\lim_{\Delta t \rightarrow 0} \frac{\eE(\X(t+\Delta
t)-\X(t)\,|\,\X(t)=\X)}{\Delta t},
\eeq
where the average $\eE(.\,|\,\X(t)=\X)$ is taken over the trajectories passing through point $\X$ at time $t$.  Similarly, the second moment is given by
 \beq \label{eq:diff}
2{\bf \Bb^{ij}}(\X)=\lim_{\Delta t \rightarrow 0} \frac{\eE( \X(t+\Delta
t)-\X(t))^i(\X(t+\Delta t)-\X(t))^j,|\X(t)=\X\rangle}{\Delta t}.
 \eeq
In practice, this inversion procedure requires combining several independent trajectories passing through each point of a domain. The drift and the diffusion tensor can be recovered from many empirical trajectories. In the next section, we generalize these formula to extract the underlying physical processes (drift and tensor) from observing a discrete ensemble of trajectories $Y_n$ at time resolution $\Delta t$.
\subsection{Recovering the drift in dimension 1}
The infinitesimal operator of the observed process $Y_n$ defined by eq. \eqref{observedp} is Gaussian and the associated stochastic discretization equation is \eqref{empi} (section \ref{ptf}). Thus, an estimator for the drift at a time resolution $\Delta $ of the observed process is
\beq \label{esta}
a_{\Delta t}(x) &=& \eE\left[\frac{Y_{n+1}-Y_n}{\Delta t} | Y_n=x\right] \\ \nonumber
&=& \frac{1}{\Delta t}\int_{\mathbb{R} }(y-x) p(Y_{n+1}=y|Y_n=x) dy \\ \nonumber
&=& a(x)+o(1).
\eeq
The average eq. \eqref{esta} computed from an observed trajectories gives the same drift component as the initial physical process in the limit
\beq
\lim_{\Delta t\rightarrow 0}a_{\Delta t}(x)=a(x).
\eeq
Thus adding a Gaussian noise on the physical process, sampled at any rate, does not alter the physical deterministic drift at first order in $\sigma$ (see appendix for the second order).
\subsection{Recovering the diffusion tensor in dimension 1}
The diffusion tensor at position $x$ of the observed trajectories is estimated as
\beq
\label{estimatordiff}
D_{\Delta t}(x) &=& E\left[\frac{(Y_{n+1}-Y_n)^2}{2\Delta t} | Y_n=x\right]\\\nonumber
&=& \frac{1}{2\Delta t}\int_{\mathbb{R} }(y-x)^2 p(Y_{n+1}=y|Y_n=x) dy \\\nonumber
&=&\frac{\sigma^2}{\Delta t}+D+\sigma^2a'(x)  +\frac{a^2(x)}{2}\Delta t +o(\Delta t),
\eeq
where the transition probability of the observed process is computed from expression \eqref{pdfosb}. This result shows that at a time resolution $\Delta t$, estimator \eqref{estimatordiff} contains an additional term $\sigma^2a'(x)$ to the diffusion coefficient of the physical process. In practice, the field $a(x)$ is recovered from estimator \eqref{esta}, and the resolution $\Delta t$ is fixed, the amplitude of the noise $\sigma$ is calibrated from instrumental noise. It is then possible to recover the diffusion coefficient $D$. A general expression for a diffusion tensor $D(x)$ is derived in appendix B.

\begin{figure}[http!]
\centering
\includegraphics[width=1\textwidth]{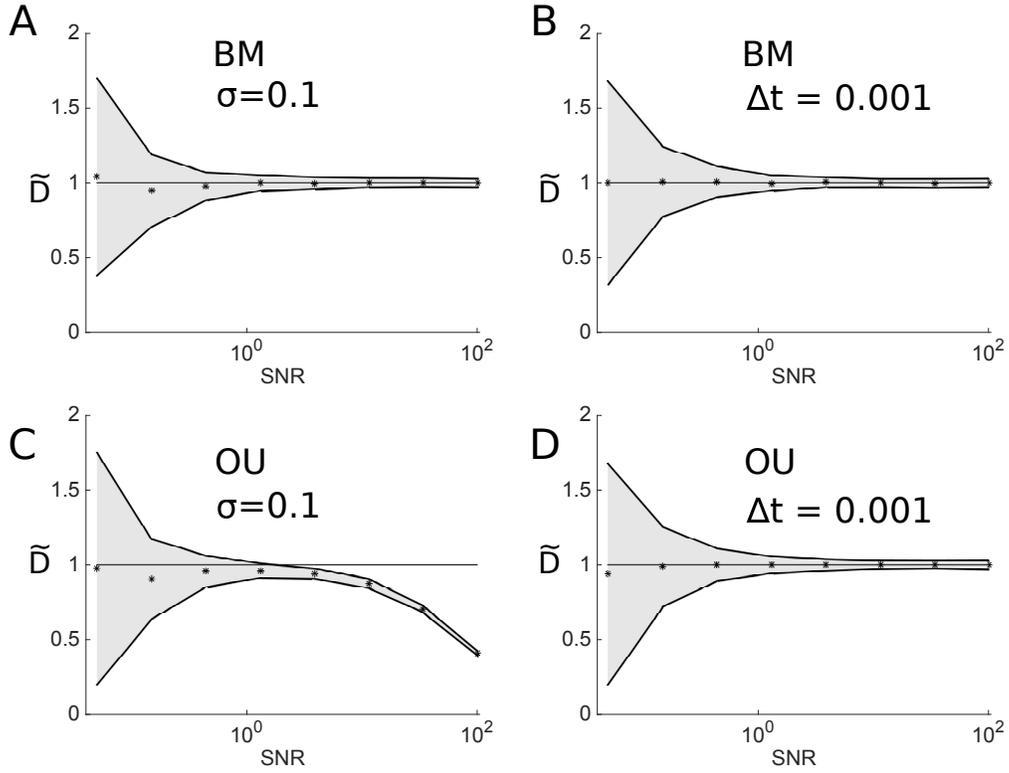}
\caption{{\bf  Estimation of the diffusion coefficient for a Brownian Motion (BM) and Ornstein-Ulhenbeck process (OU).}  Trajectories of Brownian motion (\textbf{A},\textbf{B}) and an Ornstein-Uhlenbeck process (\textbf{C},\textbf{D}), were simulated using Euler's scheme. Trajectories were sub-sampled  to 10,000 points in trajectories. A position noise was added to the process. The diffusion coefficient  $D_{\Delta t}$ is estimated using formula \eqref{diff1dmeasure}.  The signal-noise ratio (SNR) is defined as $\frac{D}{\frac{\sigma^2}{ \Delta t}}$. The dot line represents $\tilde{D} = D_{\Delta t} -\frac{\sigma^2}{\Delta t}$ and  the continuous line represents $\tilde{D} \pm std$. The diffusion coefficient is set to $D=1$ and the drift is $a(x)=-2x$. (\textbf{A,C}) The positional noise is fixed at $\sigma  =0.1$, while the sampling rate $\Delta t$ is varying.
(\textbf{B,D}) $\Delta t  =0.001$ and the position noise $\sigma$ is varying.
} \label{figTransitionMBOU}
\end{figure}

Using formula \eqref{estimatordiff}, with 10,000 points, we estimated the diffusion coefficient  $\tilde{D}$ in Fig. \ref{figTransitionMBOU}A-B. The signal to noise ratio (SNR) is defined as $\frac{D}{\frac{\sigma^2}{ \Delta t}}$. We also estimated the diffusion coefficient for an Ornstein-Uhlenbeck process (Fig. \ref{figTransitionMBOU}C-D). These numerical results show that the estimator for the diffusion coefficient is biased for a high SNR, due to the approximation $X_{n+1}-X_n \sim \mathcal{N}(a(X_n)\Delta t,\sqrt{2D\Delta t})$, which is only valid for a small time step $\Delta t$. However this approximation is acceptable for a Brownian motion, as shown in  Fig. \ref{figTransitionMBOU}A and \ref{figTransitionMBOU}B.

\subsection{Other estimators}
For a stochastic process containing a drift component, it is not possible to extract the physical diffusion coefficient directly by combining the first and the second moment estimators, which is in contrast with the pure diffusion case (see \cite{Berglund,Verstergaard}). We now present an estimator where the Gaussian instrumental noise can be eliminated. Using the difference $\Delta \Y_n=\Y_{n+1}-\Y_n$, we can rewrite
\beqq
\Delta \Y_n &=&a(\X_n) \Delta t  +{ \sigma (\X_n)}\Delta W_n+\sigma( \eta_{n+1}-\eta_n) \\
\Delta \Y_{n-1} &=& a(\X_{n-1}) \Delta t  +{ \sigma (\X_{n-1})}\Delta W_{n-1}+\sigma( \eta_{n}-\eta_{n-1}),
\eeqq
where $\Delta W_n$ and $\Delta W_{n-1}$ are two independent increments of Brownian motion. The expectation is
\beq
E\left[\frac{(Y_{n+1}-Y_n)(Y_{n}-Y_{n-1})}{\Delta t}\right]=-\frac{\sigma^2}{\Delta t} E(\eta^2_n)=-\frac{\sigma^2}{\Delta t} +o(1).
\eeq
Using relation \eqref{estimatordiff}, we obtain that
\beq
E\left[\frac{(Y_{n+1}-Y_n)^2}{2\Delta t} | Y_n=x\right] +E\left[\frac{(Y_{n+1}-Y_n)(Y_{n}-Y_{n-1})}{\Delta t}\right]=D+\sigma^2 a'(x)+o(1).
\eeq
In this estimator, the instrumental noise is averaged out. However, there are no direct procedures to get rid of the derivative of the drift term, which can be extracted from the first order moment. However, computing a derivative from noisy data should be done carefully as it introduces singularities and irregularities.
\subsection{Empirical estimators associated to an Ornstein-Uhlenbeck process}
\begin{figure}[http!]
\centering
\includegraphics[width=1\textwidth]{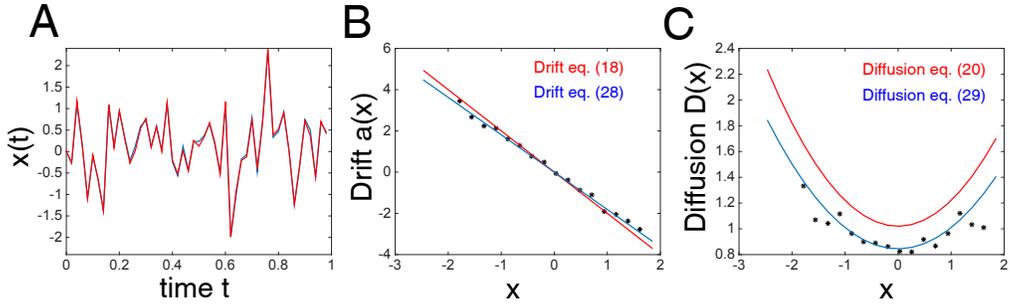}
\caption{ \textbf{Estimating the local drift and diffusion coefficient for an Ornstein-Uhlenbeck process.}  (\textbf{A})  trajectory of a one dimensional OU process. The OU process is generated using Euler's scheme (blue curve) and the observed trajectory  (red curve) is obtained by subsampling at $\Delta t = 0.1$ and with an additional position noise of standard deviation $\sigma= 0.05$ (SNR=40). The other parameters are $D=1$, $\lambda  =  2$, $\mu = 0 $. The observed trajectory contains 10,000 points. (\textbf{B}) Estimation of the local drift using eq.  \eqref{drift1dmeasure}  (black dots), and comparison
with the analytical formulas \eqref{esta} (red) and \eqref{localdriftOU} (blue). (\textbf{C}) Estimation
of the local diffusion coefficient using eq. \eqref{diff1dmeasure} (black dots), and comparison with the
analytical formulas \eqref{estimatordiff} (red) and \eqref{localdiffusionOU} (blue).}
 \label{FigCompareOU}
\end{figure}

We shall now consider an Ornstein-Uhlenbeck process,
\beq
dX  = -\lambda(X-\mu)dt + \sqrt{2D}dW,
\eeq
where the pdf is
\beq
p(y,t|x,0) = \frac{1}{\sqrt{2\pi D\frac{(1-e^{-2\lambda t})}{\lambda}}} \exp (-\frac{(y-\mu - (x-\mu)e^{-\lambda t})^2}{\frac{2D}{\lambda} (1-e^{-2\lambda t})}).
\eeq
In the discretized setting, the pdf between two time steps separated by an interval $\Delta t$ associated to the observed motion $Y_n$, can be computed from eq. \eqref{transitionprobabilityintegral} and is given by
\beq
p(Y_{n+1}=y|Y_n=x)  &=&  \intR  \intR \frac{e^{\ds-\frac{(y_1-\mu - (x_1-\mu)e^{-\lambda \Delta t})^2}{\frac{2D}{\lambda} (1-e^{-2\lambda \Delta  t})}}}{\sqrt{2\pi \frac{D}{\lambda}(1-e^{-2\lambda \Delta t})}}    \frac{e^{-\ds{\frac{(x-x_1)^2}{2\sigma^2}}}}{\sigma\sqrt{2\pi}}\frac{e^{-\ds{\frac{(y-y_1)^2}{2\sigma^2}}}}{\sigma\sqrt{2\pi}}dx_1dy_1\nonumber \\
&=& \intR  \frac{e^{-\ds{\frac{(x-x_1)^2}{2\sigma^2}}}}{\sigma\sqrt{2\pi}}  \frac{e^{-\ds{\frac{(y-\mu - (x_1-\mu)e^{-\lambda \Delta t})^2}{2(\sigma^2+ \frac{D}{\lambda}(1-e^{-2\lambda\Delta t}))}}}}{ \sqrt{2\pi(\sigma^2 +\frac{D}{\lambda}(1-e^{-2\lambda\Delta t}) )}} dx_1 \nonumber\\
&=& \frac{e^{-\ds{\frac{(y-\mu - (x-\mu)e^{-\lambda \Delta t})^2}{2(\sigma^2(1+e^{-2\lambda\Delta t})+ \frac{D}{\lambda}(1-e^{-2\lambda\Delta t}))}}}}{ \sqrt{2\pi(\sigma^2(1+e^{-2\lambda\Delta t}) +\frac{D}{\lambda}(1-e^{-2\lambda\Delta t}) )}}.
\eeq
The local dynamics can be recovered from the trajectories by computing the observed drift at time scale $\Delta t$, which  is given by
\beq
a^{OU}_{\Delta t}(x)&=& \frac{1}{\Delta t}\int_{\mathbb{R} }(y-x) p(Y_{n+1}=y|Y_n=x) dy \nonumber \\
&=& -(x-\mu)\frac{1-e^{-\lambda \Delta t}}{\Delta t},
\label{localdriftOU}
\eeq
which generalizes relation \eqref{esta}. Similarly, the observed diffusion coefficient is
\beq
D^{OU}_{\Delta t}(x)&=& \frac{1}{2\Delta t}\int_{\mathbb{R} }(y-x)^2 p(Y_{n+1}=y|Y_n=x) dy \nonumber \\
&=&  \frac{1}{2\Delta t} \left(\sigma^2(1+e^{-2\lambda\Delta t}) +\frac{D}{\lambda}(1-e^{-2\lambda\Delta t}) \right) +(\mu-x)^2 \frac{(1-e^{-\lambda \Delta t})^2}{2\Delta t}.\nonumber\\
\label{localdiffusionOU}
\eeq
In Fig. \ref{FigCompareOU}, we estimate the local drift and diffusion coefficient for an OU-process and compare the local estimators for the drift \eqref{esta} with relation \eqref{localdriftOU}. For the diffusion tensor, we compare relations \eqref{estimatordiff} and \eqref{localdiffusionOU}. At first order approximation for short time step $\Delta t$, estimators \eqref{esta}(resp. \eqref{estimatordiff}) gives similar result as \eqref{localdriftOU} (resp. \eqref{localdiffusionOU}).

\subsection{Estimating the motion of an immobile particle and criteria of detection} \label{fixedp}
When a particle is fixed at position $X_0$, the sampled trajectories are generated by the noise localization with variance $\sigma$. Computing the first moment shows that the particle is not moving and the second moment is used to extract the variance $\sigma$. The observed dynamics is given by the stochastic equation
\beqq
Y_{n}=X_0 + \sigma \eta_n,
\eeqq
where $\eta_n$ are i.i.d Gaussian variable of variance 1. The transition probability reduces to
\beqq
p(Y_{n+1}=y|Y_n=x) =  p(Y_{n+1}=y) =  \frac{e^{-\ds{\frac{(y-X_0)^2}{2\sigma^2}}}}{\sigma\sqrt{2\pi} } .
\eeqq
and the empirical estimator of the drift is
\beqq
a_{\Delta t}(x) &=&  \frac{1}{\Delta t}\int_{\mathbb{R} }(y-x) p(Y_{n+1}=y|Y_n=x) dy \\
&=& \frac{1}{\Delta t}\int_{\mathbb{R} }(y-x) \frac{e^{-\ds{\frac{(y-X_0)^2}{2\sigma^2}}}}{\sigma\sqrt{2\pi} }  dy \\
&=& -\frac{1}{\Delta t} (x-X_0),
\eeqq
which should be compared to relation \eqref{esta}: the estimator now depends on the time resolution $\Delta t$ and the location of the pinned particles, which can be determined by the empirical averaging $\frac{1}{n}\sum_{k=1}^n  Y_{k}$. The sum is converging (in probability) as $n$ goes to infinity to the mean $\eE(Y_1)=X_0$. Thus contrary to the case of a physical drift, the empirical sum $\frac{1}{N}\sum_{k}^N\left[\frac{Y^{k}_{n+1}-x}{\Delta t}\right]$ converges $-\frac{1}{\Delta t} (x-X_0)$, which depends on the time step $\Delta t$.

Similarly the second moment estimator gives for the diffusion coefficient the following expression
\beq\label{estimato2}
D_{\Delta t}(x) &=& E\left[\frac{(Y_{n+1}-Y_n)^2}{2\Delta t} | Y_n=x\right] = \frac{1}{2\Delta t}\int_{\mathbb{R} }(y-x)^2  \frac{e^{-\ds{\frac{(y-X_0)^2}{2\sigma^2}}}}{\sigma\sqrt{2\pi} } dy \nonumber \\
&=&   \frac{1}{2\Delta t} ((x-X_0)^2+\sigma^2).
\eeq
By fixing the center $x=X_0$, the empirical estimator \eqref{estimato2} allows estimating $\frac{1}{2\Delta t} \sigma^2$ and the variance $\sigma$.\\
This example is instructive because it allows differentiating a fixed particle from one trapped in a potential well (see paragraph below). In summary, the following criteria can be used: the first moment (velocity) computed from a sample trajectory for a fixed particle depends on the time resolution $\Delta t$, which is not the case for a physical particle trapped in a potential well (see relation \eqref{esta}).

\clearpage
\section{Estimators for a multidimensional diffusion process in $\ds \rR^n$}
To generalize to higher dimensions the results we derived for dimension one, we start with a $m$-dimensional stochastic equation that represents a physical processes, sampled at discrete time steps of length $\Delta t$:
\beq
\X_{n+1} =\X_{n} +\Aa (\X_n) \Delta t  + \sqrt{2D}\Delta \w
\eeq
where $\Aa$ is a vector field and $\Delta \w$ the classical $m-$dimensional centered Brownian motion of variance 1. The diffusion tensor is assumed to be a constant $D$. As described by eq. \eqref{observedp}, the observed motion is observed by the time sequences
\beq
\Y_n= \X_n+\sigma\mbox{\boldmath$\eta_n$},
\eeq
where $\mbox{\boldmath$\eta_n$}$ is a $m$-dimensional standard gaussian. The transition probability between points $\Y_n$ and $\Y_{n+1}$ is
\beqq
p(\Y_{n+1}=\Yy|\Y_n=\x)&=& \intRm  \intRm p(\X_{n+1}=\Yy_1|\X_n=\x_1)p(\Z_{n+1}=\Yy-\Yy_1) \times\\
                        &&p(\Z_n=\x-\x_1)d\x_1d\Yy_1 \\
                       &=& \intRm  \intRm p(\X_{n+1}=\Yy_1|\X_n=\x_1)
                       \frac{\ds e^{-\ds{\frac{\norm{\x-\x_1}^2}{2\sigma^2}}}}{\ds (\sigma\sqrt{2\pi})^m} \frac{\ds e^{-\ds{\frac{\norm{\Yy-\Yy_1}^2}{2\sigma^2}}}}{\ds (\sigma\sqrt{2\pi})^m}d\x_1d\Yy_1
\eeqq
Using the distribution $\X_{n+1}-\X_n \sim \mathcal{N}_m(\Aa (X_n)\Delta t, \sqrt{2D\Delta t}\IIm)$,  we obtain that the transition probability is
\beqq
p(\X_{n+1}=\Yy_1|\X_n=\x_1)  =   \frac{e^{-\ds{\frac{\norm{\Yy_1-\x_1-\Aa(\x_1)\Delta t}^2}{4D\Delta t}}}}{(\sqrt{4\pi D \Delta t})^m }.
\eeqq
We first integrate w.r.t. $\Yy_1$ and obtain
\beqq
p(\Y_{n+1}=\Yy|\Y_n=\x)= \intRm     \frac{e^{-\ds{\frac{\norm{\x-\x_1}^2}{2\sigma^2}}}}{(\sigma\sqrt{2\pi})^m}     \frac{e^{-\ds{\frac{\norm{\Yy-\x_1-\Aa(\x_1)\Delta t }^2}{4D\Delta t+2\sigma^2}}}}{\sqrt{2\pi(2 D \Delta t + \sigma^2)}^m }     d\x_1  \\
\eeqq
Changing variable $\x_1=\x+\sigma \ET$, with $\sigma \ll1$, we obtain that
 \beqq
p(\Y_{n+1}=\Yy|\Y_n=\x) =\intR  \frac{e^{-\ds{\frac{|\ET|^2}{2}}}}{(\sqrt{2\pi})^m}  \frac{e^{-\ds{\frac{(\y-\x-\sigma \ET-\Aa(\x+\sigma \ET)\Delta t)^2}{2(\sigma^2+ 2D\Delta t)}}}}{ \sqrt{2\pi(2 D \Delta t + \sigma^2)}^m} d\eta.
\eeqq
Using a Taylor expansion of the drift at the first order,
\beqq
\Aa(\x+\sigma \ET) =\Aa(\x)  + \sigma\J(\x)\ET  +o(\sigma),
\eeqq
where $\J(\x)$ is the Jacobian matrix of the vector field $\Aa$ at position $\x$.
\beqq
p(\Y_{n+1}=\Yy|\Y_n=\x) =\intR  \frac{e^{-\ds{\frac{|\ET|^2}{2}}}}{(\sqrt{2\pi})^m}  \frac{e^{-\ds{\frac{(\y-\x-\Aa(\x) \Delta t-\sigma\left( \IIm +\J(\x)\Delta t \right)\ET)^2}{2(\sigma^2+ 2D\Delta t)}}}}{ \sqrt{2\pi(2 D \Delta t + \sigma^2)}^m} d\ET.
\eeqq
%
Following the one dimensional step, from a direct integration we obtain
\beqq
p(\Y_{n+1}=\Yy|\Y_n=\x)
=  \frac{1}{\sqrt{(2\pi)^{m} \det \SSigma(\x)}}e^{\ds{-\frac{1}{2}  (  \Yy - \x -\Delta t \Aa (\x))^T  \SSigma^{-1}(\x) (  \Yy - \x -\Delta t \Aa (\x)) }}
\eeqq
where
\beq
\SSigma(\x) &=&(\sigma^2 +2D\Delta t)\IIm + \sigma^2 \B(\x) \B^T(\x)  \label{eq:B}\\
            &=&(2\sigma^2 +2D\Delta t)\IIm + \sigma^2\Delta t (\J(\x) +\J^T(\x)) +O(\Delta t^2) \nonumber
\eeq
and
\beq
\B(\x) &=& \IIm + \Delta t \J(\x).
\eeq
Formula \eqref{eq:B} generalizes to the $n$-dimensional Euclidean space the result of section \ref{ptf} for dimension one.
\subsection{Estimation of a drift and diffusion tensor}
To estimate the apparent drift and diffusion tensor, we apply analytical expressions for the pdf (formula \ref{eq:B}).
Using the formula to characterize the drift at resolution $\Delta t$ \cite{OPT} at position $\x$, we get
\beq
\aaa_{\Delta t}(\x)&=& \eE\left[\frac{\Y_{n+1}-\Y_n}{\Delta t} | \Y_n=\x\right] \nonumber\\  \nonumber
&=& \frac{1}{\Delta t}\intRm (\y-\x) p(\Y_{n+1}=\y|\Y_n=\x) d\Yy \\ \nonumber
&=&\frac{1}{(2\pi)^{m/2} \Delta t}\intRm  \frac{(\y-\x)d\Yy}{\sqrt{\det \SSigma(\x)}}e^{\ds{-\frac{1}{2}  (  \Yy-\Delta t \Aa (\x) - \x)^T  \SSigma^{-1}(\x) (  \Yy-\Delta t \Aa (\x)-\x) }}. \\ \nonumber
\eeq
Using the change of variable  $\vv =\Yy-\Delta t \Aa (\x) - \x$ we  obtain
\beq
\aaa_{\Delta t}(\x)  &=& \frac{1}{\Delta t}\intRm (\vv+\Delta t \Aa (\x))\frac{1}{\sqrt{(2\pi)^{m} \det \SSigma(\x) }}e^{\ds{-\frac{1}{2}  \vv^T  \SSigma^{-1}(\x) \vv }}   d\vv  \nonumber\\ 
&=& \Aa (\x)+o(\Delta t).
\eeq
This approximation is valid to second order in $\sigma$ (see Appendix B). Similarly in the isotropic case, the diffusion coefficient at position $\x$ can be recovered from  the second order moment approximation
\beq
D_{\Delta t}(\x) &=& E\left[\frac{\norm{\Y_{n+1}-\Y_n}^2}{2m\Delta t} | \Y_n=\x\right].
\eeq
Thus, using the pdf formula \eqref{eq:B}, we get
\beq
D_{\Delta t}(\x) &=& \frac{1}{2m\Delta t} \intRm (\vv+\Delta t \Aa (\x))^T(\vv+\Delta t \Aa (\x))\frac{1}{\sqrt{(2\pi)^{m} \det \SSigma(\x)}}e^{\ds{-\frac{1}{2}  \vv^T  \SSigma(\x)^{-1} \vv }} \nonumber   d\vv\\\nonumber
&=& \frac{1}{2m\Delta t} \intRm (\vv^T \vv + \Delta t(\Aa (\x)^T+\Aa (\x))+\Delta t^2 \Aa (\x)^T\Aa (\x))\times \nonumber\\
&&\frac{1}{\sqrt{(2\pi)^{m} \det \SSigma(\x)}}e^{\ds{-\frac{1}{2}  \vv^T  \SSigma(\x)^{-1} \vv }}    d\vv\\\nonumber
&=& \frac{1}{2m\Delta t} (Tr(\SSigma(\x)) +O( \Delta t^2)  ).
\eeq
Using eq. \eqref{eq:B}, we have
\beq
Tr(\SSigma(\x)) =m( 2\sigma^2+2D\Delta t)+2\sigma^2 \Delta  t Tr( \J(\x) ) +O(\Delta t^2).
\eeq
Finally,
\beq
D_{\Delta t}(\x)  =D+ \frac{\sigma^2}{\Delta t} + \frac{\sigma^2}{m} div(\Aa)   +O(\Delta t),
\eeq
where by definition, in local coordinates $div(\Aa)=\ds \sum_{i=1}^{m}\frac{\p a_i(\x)}{ \p x_i}$.
In general, the diffusion tensor can be approximated at order $\Delta t$ by
\beq
D^{ij}_{\Delta t}(\x) &=& E\left[\frac{(\Y_{n+1}-\Y_n)^i(\Y_{n+1}-\Y_n)^j}{2\Delta t} | \Y_n=\x\right] \\
                      &=& \frac{1}{2\Delta t} \intRm (\vv+\Delta t \Aa (\x))^i(\vv+\Delta t \Aa (\x))^j\frac{1}{\sqrt{(2\pi)^{m} \det \SSigma(\x)}}e^{\ds{-\frac{1}{2}  \vv^T  \SSigma(\x)^{-1} \vv }} \nonumber   d\vv\\
                      &=& D^{ij}+ \frac{\sigma^2}{\Delta t}\delta_{ij} + \frac{\sigma^2}{2} (\J(\x)+\J^T(\x))^{ij}   +O(\Delta t).\nonumber
\eeq


\clearpage
\section{Empirical estimators for a diffusion process using a Maximum Likelihood procedure}
We construct now parametric empirical estimators for a stochastic process using a Maximum Likelihood procedure. To recover the drift  $a(\theta_1,\dots, \theta_m)$ and diffusion coefficient we shall estimate the internal parameters $\theta_1,\dots, \theta_m$. 

We start with a sequence of observed points $(y_1,\dots,y_{N+1} )$, generated by a stochastic model 
\beq \label{eqfdt1MLE}
\dot{\x}=\As(\x,\theta_1,\thet) + \Bb(\x)\dot{\w}
\eeq
perturbed by an additive Gaussian noise, as discussed in the first section. To determine the parameters $\thet = (\theta_1,\dots, \theta_m)$, we consider the procedure which consists in maximizing the transition probability conditioned on the sequences $(y_1,\dots,y_{N+1} )$. The maximum likelihood estimator is computed from the joint probability
\beq
p( y_1,\dots,y_{N+1} , \theta_1,\dots, \theta_m).
\eeq
Assuming independent and identically distributed sample, we get 
\beq
p( y_1,\dots,y_{N+1} , \theta_1,\dots, \theta_m) = \prod_{n=1}^{N}p( y_{n+1}| y_n; \thet),
\eeq
where $p( y_{n+1}| y_n; \thet)$  is the transition probability from point $y_n$ at time $t_n$ to $y_{n+1}$ at time $t_{n}+ \Delta t$. It is given in dimension one in the entire line when $\Bb(\x)=\sqrt{2D}$ by
\beq \label{transitiontheta}
p(y_{k+1} |y_k; \thet) = \frac{e^{-\ds{\frac{\left(y_{k+1}-y_k-a(y_k,\thet )\Delta t  \right)^2}{2\sigma_{\Delta t}^2(y_k,\thet )}}}}{\sigma_{\Delta t}(y_k, \thet)\sqrt{2\pi}},
\eeq
as shown in eq. \eqref{sigma1}
\beq
\sigma_{\Delta t}(y_k, \thet) = 2\sigma^2(1+a'(y_k , \thet)\Delta t)+2D(y_k)\Delta t+O(\Delta t)^2.
\eeq
The log-likelihood is defined as
\beq\label{maxlikelihood}
\ell  (y_1,\dots,y_{N+1}   | \thet) &=&    \sum_{n=1}^{N} \log p( y_{n+1}| y_n ; \thet)\\
                                    &=& -\sum_{n=1}^N\log \sigma_{\Delta t}(y_n, \thet)
-\frac{1}{2}\sum_{n=1}^{N}\frac{(y_{n+1}-y_n-a(y_n;\thet)\Delta t)^2}{\sigma_{\Delta t}^2(y_n)}. \nonumber
\eeq
The parameters $\theta_1, \dots, \theta_m$ and $D$ are computed as maximizer of the likelihood function and thus by differentiating $\ell$ with respect to $\theta_1, \dots, \theta_m, D$. The conditions $\frac{\p \ell}{\p D}=0$ and $\frac{\p \ell}{\p \theta_i}=0$ can be written as  
\beq
\label{condition1}
\frac{\p \ell}{\p D}=0= - \sum_{n=1}^{N}\frac{\frac{\p \sigma_{\Delta t}(y_n,\thet)}{\p D}}{\sigma_{\Delta t}(y_n,\thet)} +  \sum_{n=1}^{N}\frac{\frac{\p \sigma_{\Delta t}(y_n,\thet)}{\p D}}{\sigma_{\Delta t}^3(y_n,\thet)}    (y_{n+1}-y_n -a(y_n;\thet)  \Delta t)^2.
\eeq
When the diffusion coefficient $D$ is independent of the position, the estimator is
\beq
\tilde D =\frac{1}{2\Delta t}\left(\frac{1}{N} \sum_{n=1}^{N}(y_{n+1}-y_n -a(y_n;\thet)  \Delta t)^2-2\sigma^2(1+\frac{\p a}{\p x}(y_n;\thet)\Delta t)\right)+O(\Delta t).
\eeq
Moreover, differentiation of $\ell$ w.r.t. $\theta_i$, $1\leq i < m $, gives
\beq
\frac{\p \ell}{\p \theta_i}&=&   -N\frac{\ds{\frac{\p \sigma_{\Delta t}}{\p   \theta_i}}}{\sigma_{\Delta t}} +\frac{\p \sigma_{\Delta t}}{\p   \theta_i} \sum_{n=1}^{N} \frac{(y_{n+1}-y_n-a(y_n;\thet)\Delta t)^2}{\sigma_{\Delta t}^3} \nonumber \\
&+&\frac{\Delta t}{\sigma_{\Delta t}^2} \sum_{n=1}^{N}\frac{\p a(y_n;\thet)}{ \p \theta_i}(y_{n+1}-y_n -a(y_n;\thet)\Delta t).
\eeq
Conditions  \eqref{condition1} and $\frac{\p \ell}{\p \theta_i}=0$ thus lead to the condition on the parameters $\theta_1,\dots, \theta_m$
\beq
\sum_{n=1}^{N}\frac{\p a(y_n;\thet)}{ \p \theta_i}(y_{n+1}-y_n -a(y_n;\thet)\Delta t) = 0 \mbox{ for } i=1\dots m.
\label{conditionMLE}
\eeq
\subsection{Estimating an Ornstein-Uhlenbeck process}\label{section:MLEOU1}
An Ornstein-Uhlenbeck process sampled at short time $\Delta t $ (eq. \eqref{eqsyst}) is
\beqq
X_{n+1} = X_n-\lambda( X_n -\mu)\Delta t  + \sqrt{2D}\Delta w.
\eeqq
We construct the transition probability of the observed motion as
\beq
\label{transitionOUprocess}
p(Y_{n+1}=y|Y_n=x) = \frac{e^{-\frac{(y-x+\lambda (x-\mu) \Delta t)^2}{2\sigma_{\Delta t}^2}}}{\sigma_{\Delta t}\sqrt{2\pi}},
\eeq
where
\beqq
\sigma_{\Delta t}^2=  2\sigma^2 +(2D -2\sigma^2\lambda ) \Delta t +O(\Delta t^2).
\eeqq
The log-likelihood \eqref{maxlikelihood} is now
\beqq
\ell(y_1,\dots,y_{N+1}| \lambda, D   ) = -N\log \sigma_{\Delta t} -\frac{1}{2}\sum_{n=1}^{N}\frac{(y_{n+1}-y_n+\lambda (y_n-\mu)\Delta t)^2}{\sigma_{\Delta t}^2}.
\eeqq
Conditions $\frac{\p \ell}{\p D}=0$, $\frac{\p \ell}{\p \lambda}=0$ and $\frac{\p \ell}{\p \mu}=0$ lead to
\beqq
 \sum_{n=1}^{N}y_n(y_{n+1}-y_n +\lambda (y_n-\mu)\Delta t))=0,
\eeqq
and thus the empirical estimator $\tilde \lambda $ for the parameter $\lambda $ is
\beq
\label{relationlambda1}
 \tilde \lambda = -\frac{1}{\Delta t}\frac{\sum_{n=1}^{N}y_n(y_{n+1}-y_n)}{\sum_{n=1}^{N}y_n(y_n-\tilde \mu)}.
\eeq
Similarly, using $\frac{\p \ell}{\p \mu}=0$, we obtain the condition
\beq
\label{relationlambda2}
 \tilde \mu = \frac{1}{N \tilde \lambda \Delta t}(y_{N+1}-y_1)-\frac{1}{N}\sum_{n=1}^{N}y_n.
\eeq
By combining \eqref{relationlambda1} and  \eqref{relationlambda2} we obtain
\beq
 \tilde \mu  = \ds{\frac{\sum_{n=1}^{N}y_n\sum_{n=1}^{N}y_n(y_{n+1}-y_n)-\sum_{n=1}^{N}y_n^2(y_{N+1}-y_1)} {N\sum_{n=1}^{N}y_n(y_{n+1}-y_n)-\sum_{n=1}^{N}y_n(y_{N+1}-y_1)}}.
 \eeq
Finally, using \eqref{condition1} we obtain for the diffusion coefficient the following empirical estimator
\beqq
\tilde D = \sigma^2(\tilde \lambda -\frac{1}{\Delta t}) +\frac{1}{2N\Delta t} \sum_{n=1}^{N}(y_{n+1}-y_n +  \tilde\lambda(y_n-\tilde \mu) \Delta t)^2.
\eeqq
\subsection{Estimating an Ornstein-Uhlenbeck process from the exact transition probability}\label{section:MLEOU2}
\begin{figure}[http!]
\centering
\includegraphics[width=1\textwidth]{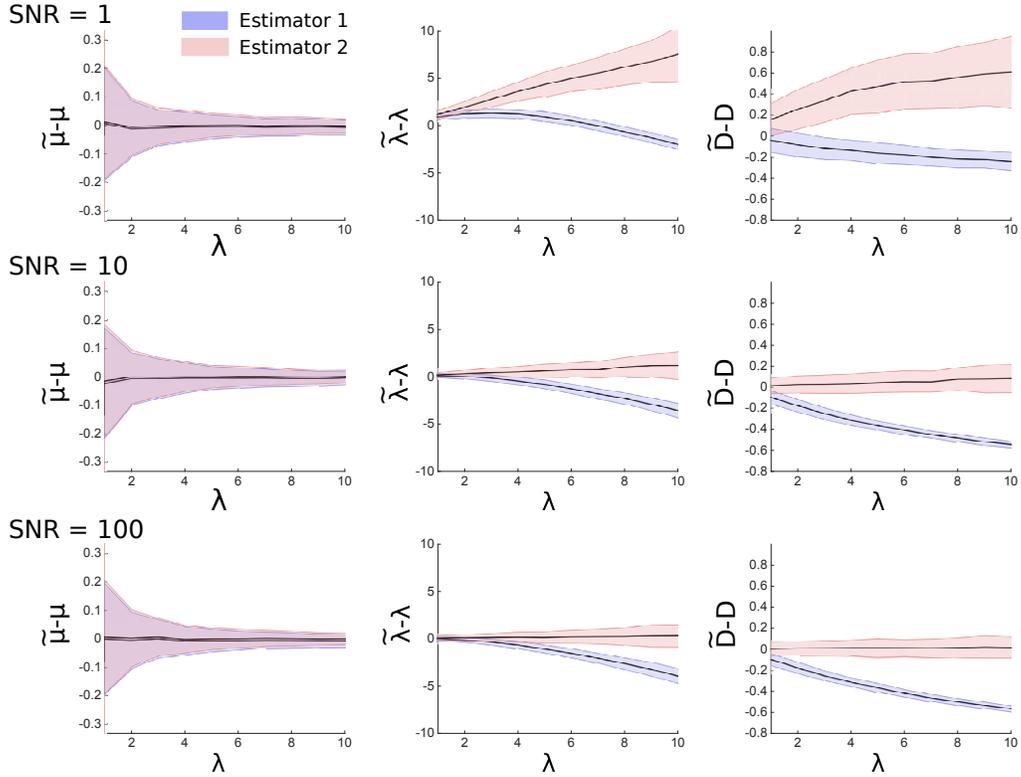}
\caption{\textbf{Comparison of the maximum-likelihood estimators of an OU process.} Comparison of the estimators in section \ref{section:MLEOU1} (blue) and \ref{section:MLEOU2} (red) for an OU process with $D=1$, $\mu= 1$, and various values of $\lambda$. From left to right, the plotted estimations are $\tilde{\mu}-\mu$,  $\tilde{\lambda}-\lambda$, and $\tilde{D}-D$. The black line is the mean of the estimation for trajectories of 500 points, and  the colored part indicates $\pm$ the standard deviation. The observation time-step is  $\Delta t$  = 0.1 and from top to bottom,  $\sigma = \sqrt{0.1} $ (SNR=1), $\sigma = \sqrt{0.01}$ (SNR=10),$\sigma = \sqrt{0.001}$ (SNR=100).} \label{MLEE}
\end{figure}

The maximum likelihood method to estimate the parameters of an  Ornstein-Uhlenbeck process uses the exact transition probability of the observed motion, given by
\beq
p(Y_{n+1}=y|Y_n=x) &=& \frac{e^{-\ds{\frac{(y-\mu - (x-\mu)e^{-\lambda \Delta t})^2}{2(\sigma^2(1+e^{-2\lambda\Delta t})+ \frac{D}{\lambda}(1-e^{-2\lambda\Delta t}))}}}}{ \sqrt{2\pi(\sigma^2(1+e^{-2\lambda\Delta t}) +\frac{D}{\lambda}(1-e^{-2\lambda\Delta t}) )}}.
\eeq
For a trajectory of $N+1$ observed points $(y_1,\dots,y_{N+1})$, the log-likelihood is
\beqq
\ell(y_1,\dots,y_{N+1}| \lambda,\mu, D   ) &=& -\frac{1}{2}N\log (\sigma^2(1+e^{-2\lambda\Delta t}) +\frac{D}{\lambda}(1-e^{-2\lambda\Delta t}) )
\\
&-&\sum_{i=1}^{N} \ds{\frac{(y_{i+1}-\mu - (y_i-\mu)e^{-\lambda \Delta t})^2}{2(\sigma^2(1+e^{-2\lambda\Delta t})+ \frac{D}{\lambda}(1-e^{-2\lambda\Delta t}))}}  .
\eeqq
Maximizing the log-likelihood leads for the parameters $ \tilde{\lambda}, \tilde{\mu}, \tilde{D}$ to the equations
\beq
 \frac{\p \ell}{\p D}(y_1,\dots,y_{N+1}|\tilde{\lambda}, \tilde{\mu}, \tilde{D}   )&=& 0\nonumber \\
\frac{\p \ell}{\p \lambda}(y_1,\dots,y_{N+1}|\tilde{\lambda}, \tilde{\mu}, \tilde{D}   )&=& 0  \nonumber \\
 \frac{\p \ell}{\p \mu}( y_1,\dots,y_{N+1}|\tilde{\lambda}, \tilde{\mu}, \tilde{D}   )&=&0.
\eeq
We are left with solving the three equations. However, because of the term $\lambda$ and in expression $e^{-\lambda \Delta t}$, it is not possible to find a closed-form solution for the parameters. In the following, we will estimate the parameter $\lambda$ using numerical optimizations. The estimator for $\mu$ and the diffusion coefficient $D$  are given by
\beq
\tilde{\mu} = \frac{1}{N} \left(\sum_{i=2}^N y_i\right)+ \frac{1}{1-e^{-\tilde{\lambda} \Delta t}}\frac{y_{N+1}-y_1 e^{-\tilde{\lambda \Delta t}}}{N},
\eeq
\beq
\tilde{D} = \frac{\tilde{\lambda}}{1-e^{-2\tilde{\lambda}\Delta t}} \left(\frac{1}{N}\sum_{i=1}^{N} (y_{i+1}-\tilde{\mu} - (y_i-\tilde{\mu})e^{-\tilde{\lambda} \Delta t})^2\right) -\sigma^2 \tilde{\lambda}\frac{1+e^{-2\tilde{\lambda}\Delta t}}{1-e^{-2\tilde{\lambda}\Delta t}}.
\eeq
{We now compare the two maximum-likelihood estimators determined in sections \ref{section:MLEOU1} and \ref{section:MLEOU2} using numerical simulations. To evaluate the performance of the estimators, we simulated trajectories following an Ornstein-Uhlenbeck process. We fixed the time-step $\Delta t=0.1$ and estimated  the parameters $\lambda, \mu $ and $D$ for $n=500$ observations. The average and standard deviation of the estimated parameters $\tilde{\lambda}, \tilde{\mu} $ and $\tilde{D}$ are obtained by taking  500 realizations of the process. Moreover, the parameters were estimated for various values of the observation noise $\sigma$. The results are summarized in Fig. \ref{MLEE}. As expected, the estimator of section  \ref{section:MLEOU2}, which is based on the actual transition probability of the OU process, gives better estimates than the estimator of section  \ref{section:MLEOU1}. }

%

\section{Discussion and Conclusion }
We presented here several empirical estimators that can be used to compute the first and second moment of a stochastic process from SPTs data. When a Gaussian noise is added to the physical process, the analysis of the estimator reveals that the drift and the diffusion tensor (formula \eqref{esta} and \eqref{estimatordiff}) are recovered at first order. 

The present estimators are very different from classical MSD, computed along trajectories. Here the estimators are based on computing {the first and second moments} using the realization of an ensemble of many trajectories. In addition, as shown in appendix B, computing the moments does not require the a priori knowledge of the probability distribution function of the process. Appendix A shows how the first two moments are computed by dividing the space in bins.

The key message of this analysis is that the drift can be recovered entirely to a first order approximation in the variance amplitude $\sigma$ (relation \eqref{eq:forceAppendix}). When the drift varies in space, the estimated diffusion tensor contains the derivative of the drift (or the divergence in higher dimension), that needs to be subtracted to recover the physical diffusion coefficient.

The present analysis provides the theoretical framework for extracting physical parameters from super-resolution single-particle trajectories \cite{HozePNAS,HozeBJ2014,Lippicott-Schwartz}, where the drift was recovered and potential wells were estimated, without accounting for the additive Gaussian external noise. Here we have shown that to a first order approximation, the additive Gaussian noise does not contribute { to the drift estimation} (only at the second order), allowing us to conclude that the estimation of the energy of the potential well was not affected (to order one) by an external localization noise. This analysis confirms that the previous results \cite{HozePNAS,HozeBJ2014,Lippicott-Schwartz} are valid even if there is a Gaussian empirical noise added.

Finally, another key result here is that physical potential wells cannot be mixed with a fixed particle, as sampled with a noisy measurement. We provided here a criteria to characterize that a particle is fixed and not confined (section \ref{fixedp}). In particular, detected converging arrows in a vector field \cite{HozePNAS,HozeBJ2014} extracted from SPT analysis reveals a physical potential well and cannot have been generated by fixed particles due to the tracking algorithm. This is even more evident when wells are not isotropic. However the present analysis does not reveal the origin of the wells.


\section{Appendices}
\section*{Appendix A: Approximation formula for the local drift and diffusion coefficient}
Computations with the estimators developed here from empirical data depend on the following steps: starting with a sample of $N_t$ observed trajectories $\{y^i(t_j),\,i=1,2,\ldots N_t,\,j=1,2,\ldots,N_s\}$, where $t_j$ are the sampling times, and $N_s$ is the number of points in each trajectory, the dynamics is reconstructed by computing the local drift and diffusion coefficient of the observed diffusion process.
First, the range of points on the line is partitioned into $M$ bins of width $r$, centered at $x_k$, such that
$$x_1-\frac{r}{2}<\min\{y^i(t_j),\,1\leq i\leq N_t,\, 1\leq j\leq N_s\}$$ and
$$x_M+\frac{r}{2}>\max\{y^i(t_j),\,1\leq i\leq N_t,\, 1\leq j\leq N_s\}.$$
The effective  drift and  diffusion coefficient of the observed diffusion process are evaluated in each bin from the empirical versions of the formulas \cite{DSP,Karlin}
\beq
a_{\Delta t}(x) =&\, \underset{\Delta t\rightarrow 0}{\lim} \frac{1}{ \Delta
t}\eE\left[y(t+\Delta t)-y(t)\,|\,y(t)= x\right]\label{Ea}\\
2D_{\Delta t}(x)=&\,\lim_{\Delta t \rightarrow 0}
\frac{1}{ \Delta t}\eE\left[\left[y(t+\Delta t)-y(t)\right]^2\,|\,y(t)=x \right].\label{ED}
\eeq
The empirical version of (\ref{Ea}) at each bin point $x_k$ is
\beq
a_{\Delta t}(x_k) = \frac{1}{N_k} \sum_{i=1}^{N_t}\sum_{j=1, y^i(t_j)\in B\left(x_k,\Delta
x \right)}^{N_s}\frac{y^i(t_{j+1})-y^i(t_j)}{\Delta t},\label{drift1dmeasure}
\eeq
where $B(x_k,r)$ is the bin $\left[ x_k-r/2, x_k +r/2 \right]$. The condition $y^i(t_j)\in B\left(x_k,r \right)$ in
the summation means that $ |y^i(t_j)-y_k|<r/2$. The points $y^i(t_j)$ and $y^i(t_{j+1})$ are sampled consecutively from the $i$th
trajectory such that $y^i(t_j)\in B(x_k,r)$ {and the} number of points in $B(x_k,r)$ is $N_k$. Similarly, the empirical version of (\ref{ED}) at bin point  $x_k$ is
\beq
D_{\Delta t}(x_k) = \frac{1}{N_k} \sum_{j=1}^{N_t}\sum_{j=1, \tilde y^i(t_j) \in B(x_k,r)}^{N_s}\frac{\left[y^i(t_{j+1})-y^i(t_j)\right]^2}{2\Delta t}.
\label{diff1dmeasure}
\eeq
\section*{Appendix B: higher order moment estimates and general inversion formula}
We present now a different approach to estimate the drift and diffusion coefficients by using direct regular expansion. This approach do not assume any knowledge of the pdf of the process and is thus applicable to any general manifold. We start with the continuous stochastic equation of  \eqref{eqsyst}
\beq
\dot{\X} =\As(\X) + \Bb(\X)\dot{\w},
\eeq
and
\beq
\dot{\Y}= \dot{\X}+\sigma \dot{\ET},
\eeq
where both $\w$ and $\ET$ are two independent identically distributed Brownian variables.
The close stochastic equation for $\Y$ is
\beq
\dot{\Y} =\As(\Y- \sigma\ET) + \Bb(\Y- \sigma\ET)\dot{\w}+\sigma \dot{\ET},
\eeq
Using a Taylor expansion to order $k$, we get
\beq
\As(\Y- \sigma\ET)&=& \sum_{0}^k  \frac{(-\sigma)^k}{k!}\frac{\p \As^{k}(\Y)}{\p\x^k}\ET^k +O(\sigma^{k+1}), \\
\Bb(\Y- \sigma\ET)&=& \sum_{0}^k  \frac{(-\sigma)^k}{k!}\frac{\p \Bb^{k}(\Y)}{\p\x^k}\ET^k +O(\sigma^{k+1}).
\eeq
Using a second order expansion we obtain that in dimension one,
\beq
\dot{Y} =a(Y)- \sigma\ET a'(Y)+ \frac{\sigma^2}{2}\ET^2a''(Y) + (\Bb(Y)- \sigma\ET\Bb'(Y)+ \frac{\sigma^2}{2}\ET^2\Bb''(Y))\dot{w}+\sigma \dot{\ET}.
\eeq
Thus, the expectation is
\beq \label{eq:forceAppendix}
\lim_{\Delta t\rightarrow 0}\frac{\eE_{w,\ET}(\Y(t+\Delta t)-\Y(t)\,|\,\Y(t)= Y)}{\Delta t}&=&a(Y)+\frac{\sigma^2}{2}\eE_{\ET}(\ET^2) a''(Y)+o(\sigma^2) , \nonumber \\
&=& a(Y)+\frac{\sigma^2}{2} a''(Y)+o(\sigma^2),
\eeq
where we used that $\eE_{\ET}(\ET^2)=1$. We conclude that at order two, a correction has to be added to the drift, but when $\sigma$ is small this contribution is negligible. In particular, this result shows that at the first order the additive noise do not influence the recovery of the vector field and local potential wells. The energy is thus not affected by this additive noise.

Similarly, the diffusion coefficient is computed from the second moment
\beq
\frac{\eE_{w,\ET}((\Y(t+\Delta t)-\Y(t))^2\,|\,\Y(t)=\Y)}{2\Delta t}&=& \frac{1}{2}b^2(Y)+ \sigma^2 {a'(Y)}+\frac{\sigma^2}{2\Delta t}\nonumber \\
&+&\frac{1}{2}\sigma^2 (b'^2(Y)+\frac{ b(Y)b''(Y)}{2}) \nonumber\\
&+&o(\Delta t)+o(\sigma^2).
\eeq
The analysis presented here can be generalized to $n$-dimension and does not depend on any a priori information about the pdf of the stochastic process to be estimated.

\section*{Acknowledgment}
We thank Suliana Manley for insightful discussions.


\end{document}